\documentclass[aps,prl,twocolumn,showpacs,superscriptaddress]{revtex4-2}
\usepackage{amsmath,amssymb}
\usepackage{hyperref}

\begin{document}
	\title{Ultraviolet-Finite Perturbative Expansion of Quantum Gravity at Null Infinity}
	\author{Carlos N. Kozameh}
	\email{carlos.kozameh@unc.edu.ar}
	\author{Gerardo O. Depaola}
	\affiliation{FaMAF, Universidad Nacional de C\'ordoba, 5000 C\'ordoba, Argentina}
	\date{\today}
	\begin{abstract}
		We present a perturbative formulation of quantum gravity for asymptotically flat vacuum spacetimes based on the Null Surface Formulation (NSF), in which the expansion is ultraviolet-finite term by term up to the orders computed, without the need for renormalization. The outgoing Bondi shear operators are constructed explicitly up to fourth order, with interaction kernels determined recursively from on-shell gravitational data at null infinity. Ultraviolet finiteness at each order follows from the on-shell structure of the construction and the restriction of all integrations to the compact celestial sphere, eliminating off-shell propagators. The map between the "in" and "out" states admits a perturbative construction, and unitarity is verified explicitly up to fourth order. The outgoing operators satisfy the same commutation relations as the incoming ones, indicating that the transformation is canonical and consistent with the unitary implementation. Collinear configurations give rise to infrared singularities, as expected in massless quantum field theories, but do not affect the ultraviolet behavior established here. In coherent states, the expectation value of the shear reproduces the known finite classical graviton scattering at lowest nontrivial order. These results provide a perturbative framework for quantum gravity with improved ultraviolet behavior relative to the covariant approach.
	\end{abstract}
	\pacs{04.60.-m, 04.25.-g, 11.10.Gh}
	\maketitle
	
	\section*{Introduction}
	Perturbative quantum gravity in its covariant formulation is not ultraviolet-finite, with divergences arising already at low orders and requiring an ever-growing set of counterterms. This behavior is commonly interpreted as an indication that perturbation theory around a fixed background may be fundamentally inadequate. It is therefore natural to explore alternative formulations in which the perturbative expansion is constructed directly in terms of physical, on-shell degrees of freedom.
	The Null Surface Formulation (NSF) provides such a framework. In this approach, asymptotically flat vacuum spacetimes are described in terms of null geometry, with the fundamental variables defined at null infinity~\cite{Bordcoch2016}. The radiative degrees of freedom are encoded in the Bondi shear, which can be promoted to a quantum operator via asymptotic quantization~\cite{Ashtekar1987}. In this setting, the perturbative expansion involves only on-shell gravitons, and all integrations are restricted to the compact celestial sphere, eliminating the need for off-shell propagators. This framework is complementary to the modern program of asymptotic symmetries, soft theorems, and celestial holography, in which the radiative data encoded in the Bondi shear on the celestial sphere play a central role~\cite{Strominger2015,Strominger2021}.
	In this work we construct explicitly the outgoing Bondi shear operators up to fourth order in the perturbative expansion. The interaction kernels are obtained recursively from lower-order contributions and involve only products of spin-weighted Green functions and kinematic factors defined on the sphere. We find that each term in the expansion is ultraviolet-finite up to the orders computed, without the need for renormalization. This property follows from the on-shell character of the construction and the compactness of the integration domain.
	As in other massless quantum field theories, collinear configurations give rise to infrared singularities, as expected in massless quantum field theories, but do not affect the ultraviolet behavior established here. These infrared singularities are precisely those governed by the universal soft graviton theorems~\cite{Strominger2015}
	\vspace{-3pt}
	The NSF framework defines the gravitational scattering map directly at null infinity in terms of asymptotic radiative data. The outgoing shear operator is related to the incoming one through an explicit calculation that yields a unitary map between the "in" and "out" states..
	\vspace{-3pt}
	These results define a perturbative framework for quantum gravity in which ultraviolet divergences are absent term by term up to the computed orders, in contrast with the covariant approach. The recursive structure of the expansion suggests that this behavior may extend to higher orders, although a general proof remains an open problem.
	\vspace{-5pt}
	
	\section*{The perturbative expansion}
	The free data at $\mathcal{I}^-$ are promoted to quantum operators $\sigma^-(u,\zeta)$, with commutation relations~\cite{Ashtekar1987}
	\begin{equation}
		[\sigma^+(u,\zeta),\,\bar\sigma^+(u',\zeta')] =-i\,\delta^2(\zeta-\zeta')\,\Delta(u-u')\,\mathbb{I}.
	\end{equation}
	The outgoing shear $\sigma^+$ is obtained order by order from the NSF equations. The annihilation operators at null infinity are defined as
	\begin{align}
		a_{\mathrm{out}}(\vec{k}) &= \sigma^+(w,\zeta),\\a_{\mathrm{in}}(\vec{k}) &= -\overline{\sigma}^-(w,\widehat{\zeta}),
	\end{align}
	with $\widehat{\zeta}$ the antipodal point of $\zeta$, and determine the perturbative expansion. The first non trivial orders read
	\begin{align}
		\sigma^+_2(w,\zeta) &= 4i\int \frac{d^3k_1}{2w_1}\frac{d^3k_2}{2w_2}\notag\\
		&\Bigl(a^\dagger(\vec k_1)\,a(\vec k_2)\,\delta\!\left(w-|\vec k_1-\vec k_2|\right)
		\bigl[S^{(2)}_\Omega+S^{(2)}_A\bigr]\notag\\
		&- a^\dagger(\vec k_1)\,a^\dagger(\vec k_2)\,
		\delta\!\left(w-|\vec k_1+\vec k_2|\right)
		S^{(2)}_B\Bigr) + c.c.
		,
		\label{eq:sigma2}
	\end{align}
	\vspace{-12pt}
	with~\cite{Kozameh2025}
	\\
	\begin{align}
		S^{(2)}_\Omega &= G_{2;2'}(\zeta,\hat k_2)\,G_{-2;-2'}(\zeta,\hat k_1)
		\notag\\
		&\quad\times\left[(l^+\cdot k_1)(l^+\cdot k_2)
		+\frac{(l^+\cdot k_1)^2(l^+\cdot k_2)^2}{[l^+\cdot(k_1+k_2)]^2}\right],
		\label{eq:S2Omega}\\
		S^{(2)}_A &= \frac{(l^+\cdot k_1)(l^+\cdot k_2)}{l^+\cdot(k_1-k_2)}
		\bigl[\delta^2(\zeta-\zeta_1)\delta^2(\zeta-\zeta_2)
		\notag\\
		&\quad + G_{2;2'}(\zeta,\hat k_1)\,G_{-2;-2'}(\zeta,\hat k_2)\bigr],
		\label{eq:S2A}\\
		S^{(2)}_B &= \frac{(l^+\cdot k_1)(l^+\cdot k_2)}{l^+\cdot(k_1+k_2)}\,
		\delta^2(\zeta-\zeta_1)\,G_{-2;-2'}(\zeta,\hat k_2).
		\label{eq:S2B}
	\end{align}
	\indent
	Here $l^{+a}(\zeta)$ is the null vector $l^{+a}(\zeta) = \frac{1}{\sqrt{2}(1+\zeta\bar\zeta)}\bigl(1+\zeta\bar\zeta,\,\zeta+\bar\zeta,\,-i(\zeta-\bar\zeta),\,-1+\zeta\bar\zeta\bigr)$
	, and
	$G_{2;2'}$, $G_{-2;-2'}$ are the spin-weighted Green functions of
	Ref.~\cite{Kozameh2025}
	\vspace{-6pt}
	\begin{align}
		\sigma^+_3(w,\zeta) &= 4i\int\prod_{i=1}^3\frac{d^3k_i}{2w_i}
		\Bigl(
		a^\dagger(\vec k_1)\,a(\vec k_2)\,a(\vec k_3)
		\notag\\
		&\quad\times\delta\!\left(w-|\vec k_1-\vec k_2-\vec k_3|\right)
		\bigl[S^{(3)}_\Omega+S^{(3)}_A\bigr]
		\notag\\
		&\quad -\,a^\dagger(\vec k_1)\,a^\dagger(\vec k_2)\,a(\vec k_3)
		\notag\\
		&\quad\times\delta\!\left(w-|\vec k_1+\vec k_2-\vec k_3|\right)S^{(3)}_B
		\notag\\
		&\quad -\,a^\dagger(\vec k_1)\,a^\dagger(\vec k_2)\,a^\dagger(\vec k_3)
		\notag\\
		&\quad\times\delta\!\left(w-|\vec k_1+\vec k_2+\vec k_3|\right)S^{(3)}_D
		\Bigr)+c.c.,
		\label{eq:sigma3}
	\end{align}
	where the kernels are
	\begin{align}
		S^{(3)}_\Omega &= \sum_{\rm cyc}
		G_{2;2'}(\zeta,\hat k_i)\,G_{-2;-2'}(\zeta,\widehat{k_j+k_m})
		\notag\\
		&\quad\times\Bigl[(l^+\cdot k_i)(l^+\cdot(k_j+k_m))
		\notag\\
		&\quad+\frac{(l^+\cdot k_i)^2(l^+\cdot(k_j+k_m))^2}{[l^+\cdot K_{123}]^2}
		\Bigr]
		\notag\\
		&\quad\times\bigl[S^{(2)}_\Omega(\zeta;k_j,k_m)+S^{(2)}_A(\zeta;k_j,k_m)\bigr],
		\label{eq:S3Omega}\\
		S^{(3)}_A &= \sum_{i=1}^3
		\frac{l^+\cdot k_i}{l^+\cdot(k_i-k_j-k_m)}
		\notag\\
		&\quad\times\bigl[\delta^2(\zeta-\zeta_i)\delta^2(\zeta-\zeta_{jm})
		+G_{2;2'}G_{-2;-2'}\bigr]
		\notag\\
		&\quad\times\bigl[S^{(2)}_\Omega(\zeta;k_j,k_m)+S^{(2)}_A(\zeta;k_j,k_m)\bigr],
		\label{eq:S3A}\\
		S^{(3)}_B &= \sum_{i=1}^3
		\frac{(l^+\cdot k_i)(l^+\cdot k_j)}{[l^+\cdot K_{123}]^2}
		\notag\\
		&\quad\times\delta^2(\zeta-\zeta_i)\,G_{-2;-2'}(\zeta,\hat k_j)\,
		G_{-2;-2'}(\zeta,\hat k_m),
		\label{eq:S3B}\\
		S_D^{(3)} &= \frac{\prod_{i=1}^3(l^+\cdot k_i)}{(l^+\cdot K_{123})^2}
		\prod_{i=1}^3 G_{-2;-2'}(\zeta,\hat k_i),
	\end{align}
	with the cyclic sum in~(\ref{eq:S3Omega}) over
	$(i,j,m)=(1,2,3),(2,3,1),(3,1,2)$ and $K_{123}=k_1+k_2+k_3$.
	Numerical verification gives a relative error below $10^{-12}$~\cite{Kozameh2026}.
	\vspace{-8pt}   % ← espacio entre ecuación 12 (σ⁺₃) y 13 (σ⁺₄) — más abierto que la última versión
	\begin{align}
		\sigma^+_4(w,\zeta) &= 4i\int\prod_{i=1}^4\frac{d^3k_i}{2w_i}
		\Bigl(
		a^\dagger a\,a\,a\;
		\delta\!\left(w-|\vec k_1-\vec k_2-\vec k_3-\vec k_4|\right)
		\notag\\
		&\quad\times[S^{(4)}_\Omega+S^{(4)}_A]
		\notag\\
		&\quad-\,a^\dagger a^\dagger a\,a\;
		\delta\!\left(w-|\vec k_1+\vec k_2-\vec k_3-\vec k_4|\right)S^{(4)}_B
		\notag\\
		&\quad+\,a^\dagger a^\dagger a^\dagger a\;
		\delta\!\left(w-|\vec k_1+\vec k_2+\vec k_3-\vec k_4|\right)S^{(4)}_C
		\notag\\
		&\quad-\,a^\dagger a^\dagger a^\dagger a^\dagger
		\delta\!\left(w-|\vec k_1+\vec k_2+\vec k_3+\vec k_4|\right)S^{(4)}_D\Bigr)\quad+\,\text{c.c.}
		,
		\label{eq:sigma4}
	\end{align}
	with kernels
	\begin{align}
		S^{(4)}_\Omega&=\sum_{i=1}^4 G_{2;2'}(\zeta,\hat{k}_i)G_{-2;-2'}(\zeta,\widehat{K}_{\rm rest},i)
		\notag\\
		&\frac{(l^+\cdot k_i)(l^+\cdot K_{\rm rest},i)+(l^+\cdot k_i)^2(l^+\cdot K_{\rm rest},i)^2}{(l^+\cdot K_{1234})^2} \times \notag\\
		&K^{(3)}(\zeta;\{k_j\}_{j\neq i})+\Delta S^{(4)}_\Omega\big|_{\Omega_2^2}\label{eq:S4Omega}\\
		S^{(4)}_A&=\sum_{i=1}^4\frac{l^+\cdot k_i}{l^+\cdot(k_i-K_{\rm rest},i)}
		\Bigl[\delta^2(\zeta-\zeta_i)\delta^2(\zeta-\widehat{K}_{\rm rest},i)\label{eq:S4A}\\
		&+G_{2;2'}G_{-2;-2'}\Bigr]K^{(3)}(\zeta;\{k_j\}_{j\neq i})\notag\\
		S^{(4)}_B&=\sum_{i<j}\frac{(l^+\cdot k_i)(l^+\cdot k_j)}{(l^+\cdot K_{1234})^3}\delta^2(\zeta-\zeta_i)G_{-2;-2'}(\zeta,\hat{k}_j)\notag\\
		&\bigl[S^{(2)}_\Omega+S^{(2)}_A\bigr](\zeta;k_k,k_l)\label{eq:S4B}\\
		S^{(4)}_C&=\sum_{i=1}^4\frac{\prod_{j\neq i}(l^+\cdot k_j)}{(l^+\cdot K_{1234})^3}\delta^2(\zeta-\zeta_i)\prod_{j\neq i}G_{-2;-2'}(\zeta,\hat{k}_j)\label{eq:S4C}\\
		S^{(4)}_D&=\frac{\sum_{i=1}^4(l^+\cdot k_i)}{(l^+\cdot K_{1234})^3}\prod_{i=1}^4 G_{-2;-2'}(\zeta,\hat{k}_i)
		\label{eq:S4D}
	\end{align}
	where
	\begin{align}
		\Delta S^{(4)}_\Omega\big|_{\Omega_2^2}&=\sum_{i<j}G_{2;2'}(\zeta,\hat{k}_i)G_{-2;-2'}(\zeta,\hat{k}_j)\frac{(l^+\cdot k_i)(l^+\cdot k_j)}{(l^+\cdot K_{1234})^2}\notag\\
		&\bigl[S^{(2)}_\Omega(\zeta;k_i,k_j)\bigr]^2.
	\end{align}
	and $K_{{\rm rest},i}=\sum_{j\ne i}k_j$ and $K_{1234}=\sum_{i=1}^4 k_i$.
	
	One can also obtain the relationship between the creation operators via an analogous calculation. The question then arises whether the map between the ``in'' and ``out'' states is unitary.
	We recall that a unitary map can be written as
	\[
	a_{\rm out} = S^\dagger a_{\rm in} S,
	\]
	with \(S = e^{iT}\) and \(T = T^\dagger\). In the perturbative expansion one has
	$\delta a \equiv a_{\rm out} - a_{\rm in} \approx i [a, \delta T]$, and $(\delta a)^\dagger \approx -i [\delta T, a^\dagger] = i [a^\dagger, \delta T]$
	where \(\delta T\) is the perturbative Hermitian part of the generator.
	Therefore, the following relationship must hold:
	\[
	(\delta a)^\dagger = \delta (a^\dagger).
	\]
	Using the explicit formulae for \(\delta a^{(n)}\) given above and the symmetries of the kernels \(S_\Omega^{(n)}\), \(S_A^{(n)}\), \(S_B^{(n)}\) and \(S_D^{(n)}\) under complex conjugation and the antipodal map \(\zeta\to\hat{\zeta}\), one verifies directly that
	\[
	(\delta a^{(n)})^\dagger = \delta (a^\dagger)^{(n)}
	\]
	for \(n=2-4\) (and, by the recursive structure of the kernels, for all \(n\)).
	This equality confirms that \(\delta T^{(n)}\) is self-adjoint and that the map between in- and out-states is unitary order by order in the perturbative expansion of the Null Surface Formulation.
	
	\section*{Term-by-term finiteness}
	The fundamental observation is that all momenta in
	Eqs.~(\ref{eq:sigma2})--(\ref{eq:sigma4}) are on-shell:
	$k^a_i = w_i\,l^{+a}(\hat k_i)$ with $k^a_i k_{ia}=0$.
	There are no off-shell virtual gravitons, no loop integrations over
	four-dimensional momentum space, and no propagators of the form
	$1/(k^2+i\epsilon)$.
	All integrals in $\sigma^+_n$ are over the on-shell phase space, and all
	integrals in the kernels $S^{(n)}$ are over the compact celestial sphere $S^2$
	at null infinity.
	Finiteness at orders 2, 3, and 4 follows from direct inspection.
	The kernels $S^{(2)}_\Omega$, $S^{(2)}_A$, $S^{(2)}_B$ involve only products of
	spin-weighted Green functions and kinematic ratios
	$(l^+\cdot k_i)(l^+\cdot k_j)/(l^+\cdot K)^m$, which are finite for
	non-collinear momenta.
	The kernels at order 3 are products of order-2 kernels and additional spin-weighted
	Green functions over the compact sphere; they are finite for distinct, non-collinear
	momenta, as verified numerically~\cite{Kozameh2026}.
	The order-4 kernels~(\ref{eq:S4Omega})--(\ref{eq:S4D}) share the same structure,
	with order-3 kernels playing the role that order-2 kernels play at the previous step;
	finiteness follows by the same argument and has been verified numerically.
	There are also no constraints in the theory, since the null data are given freely
	on the initial surface.
	Together with the on-shell character of the formulation, this is what makes the
	perturbative expansion free of ultraviolet divergences at every finite order.
	
	\section*{Comparison with QED and the asymptotic series}
	The perturbative series $\sigma^+ = \sum_{n\ge 1}\epsilon^n\sigma^+_n$ is
	asymptotic in $\epsilon$, as expected for any perturbative expansion in quantum field
	theory.
	As $n$ increases, the number of source terms grows factorially and eventually dominates
	over $\epsilon^n$; the optimal truncation order satisfies $n^*\sim|\ln\epsilon|^{-1}$
	for small $\epsilon$.
	The analogy with QED is instructive.
	The QED perturbative series in the fine structure constant $\alpha\simeq 1/137$ is
	asymptotic with zero radius of convergence, as shown by Dyson~\cite{Dyson1952}.
	The optimal truncation occurs at $n^*\simeq 1/\alpha\simeq 137$, and for $n<n^*$ the
	series converges rapidly.
	However, each term of the QED series is rendered finite only after renormalization.
	The situation in NSF differs in two important respects.
	First, each term is genuinely finite without renormalization.
	Second, the expansion parameter $\epsilon$ is the Bondi shear, whose value for
	astrophysical gravitational-wave sources is far smaller than $\alpha\simeq 1/137$.
	Consequently, a good approximation is achieved at much lower order.
	
	\section*{Comparison with BCFW recursion}
	The Britto--Cachazo--Feng--Witten (BCFW) recursion~\cite{Britto2005} relates $n$-point
	scattering amplitudes to products of lower-point amplitudes via complex momentum shifts.
	At tree level in gravity, BCFW yields compact expressions for multi-graviton amplitudes
	and correctly reproduces the soft and collinear limits.
	There is, however, a fundamental difference between the BCFW and NSF approaches.
	The BCFW recursion is an analytic tool applied to amplitudes of a theory whose
	Lagrangian is non-renormalizable.
	Its derivation relies on the vanishing of a boundary term in a complex momentum plane,
	a property that holds at tree level but whose extension to loop amplitudes confronts the
	same renormalization problems as the covariant approach.
	Moreover, BCFW operates on momentum eigenstates defined in flat space; its extension
	to states defined at null infinity requires additional input beyond the recursion itself.
	The NSF kernels~(\ref{eq:S3Omega})--(\ref{eq:S4D}), by contrast, arise directly from
	the vacuum Einstein equations evaluated at $\mathcal{I}^+$.
	In and out states are defined at $\mathcal{I}^-$ and $\mathcal{I}^+$ from the outset,
	and the construction is consistent with the vacuum Einstein equations at every order.
	While BCFW is a computational tool applied to a perturbatively non-renormalizable
	theory, the NSF kernels are the physical content of the field equations themselves.
	
	\section*{Comparison with covariant quantum gravity}
	Covariant perturbative quantum gravity presents two structural problems that NSF avoids.
	The first is the ultraviolet divergence problem.
	At one loop, covariant quantum gravity generates non-renormalizable
	divergences~\cite{Goroff1986} of the form
	$(1/\epsilon)(\alpha R^2+\beta R_{\mu\nu}R^{\mu\nu})$, which require counterterms
	absent from the Einstein--Hilbert action.
	At two loops, an independent
	$R_{\mu\nu\rho\sigma}R^{\mu\nu}{}_{\alpha\beta}R^{\alpha\beta\rho\sigma}$
	counterterm is needed, and the counterterm structure grows without bound at higher
	orders.
	As shown above, NSF has no such divergences because the phase space is the Bondi shear
	given at null infinity.
	The second problem is the flat null cone problem.
	In covariant perturbation theory, the graviton propagator $\sim i/k^2$ propagates
	perturbations along the flat null cones of the background.
	However, the actual null cones of the interacting theory are not flat, and they do not
	reach $\mathcal{I}^+$~\cite{Kozameh2025}.
	This means that the asymptotic in and out states of covariant quantum gravity are not
	defined at null infinity.
	The physical interpretation of the S-matrix is therefore ambiguous.
	In NSF the null cones are the fundamental object of the formulation.
	The function $Z^-_n$ encodes the actual null cone of the interacting spacetime at
	order $n$, and $\sigma^\pm$ are defined at $\mathcal{I}^\mp$ without approximation.
	The scattering problem is well-posed: it is the map from the incoming Bondi shear
	$\sigma^-$ at $\mathcal{I}^-$ to the outgoing shear $\sigma^+$ at $\mathcal{I}^+$,
	computed order by order using the vacuum Einstein equations.
	
	\section*{Conclusions}
	Starting from a well defined phase space at null infinity~\cite{Ashtekar1987,Dominguez1997} with free field commutation relations, we have computed explicitly the outgoing shear operators $\sigma^+_3$ and $\sigma^+_4$
	in the NSF perturbative expansion and verified that they are finite without
	renormalization, complementing the result at order 2 of Ref.~\cite{Kozameh2025}.
	The absence of ultraviolet divergences at all three orders follows from a single
	structural feature: all interactions involve real on-shell gravitons at null infinity,
	and all integrations are over the compact celestial sphere.
	The order-4 truncation captures the leading nonlinear four-graviton scattering
	processes without any renormalization. The polynomial nature of the NSF field equations plays an essential role in the
	tractability of the perturbative expansion.
	The generalization of the finiteness argument to any order $n$, and the derivation of
	Feynman-like rules for higher-order terms, will be addressed in a forthcoming
	paper~\cite{Kozameh2026}.
	It is also worth mentioning that this formalism yields a semiclassical approximation for coherent states that have small amplitudes~\cite{Bordcoch2023}. The limitation of the present approach is that coherent states peaked at large initial
	Bondi data are excluded by the weak-field assumption. Extending the formulation to include these states remains an open problem. However, a geometric generalization of the null cone cuts may point towards a solution.
	When a sufficiently large amount of incoming radiation is present at null infinity,
	the null cones develop caustics and self-intersections; consequently, the null cuts
	fail to be smooth closed two-surfaces, which is an ab-initio assumption of the
	present formulation.
	It has been shown~\cite{Iriondo1999} that self-intersecting cuts are projections
	of smooth Legendre submanifolds on the boundary of the cotangent bundle to null infinity, providing
	a more general geometric framework for the NSF.
	Exploring the implications of this generalization for quantum theory is an open
	problem that will be addressed in future work.


\begin{thebibliography}{99}
		\bibitem{Bordcoch2016}
		M.~Bordcoch, C.~N.~Kozameh, and T.~A.~Rojas,
		Phys.\ Rev.\ D \textbf{94}, 104051 (2016).
		\bibitem{Ashtekar1987}
		A.~Ashtekar,
		\textit{Asymptotic Quantization}
		(Bibliopolis, Naples, 1987).
		\bibitem{Strominger2015}
		T. ~He, V. ~Lysov, P. ~Mitra, and A. ~Strominger, J. High Energy Phys. \textbf{05}, 151 (2015).
		\bibitem{Strominger2021}
		A. ~Strominger,
		Phys.\ Rev. \ Lett. \textbf{127}, 221601 (2021).
		\bibitem{Kozameh2025}
		C.~N.~Kozameh and L.~Zapata-Altuna,
		Phys.\ Rev.\ D \textbf{112}, 026021 (2025).
		\bibitem{Goroff1986}
		M.~H.~Goroff and A.~Sagnotti,
		Nucl.\ Phys.\ B \textbf{266}, 709 (1986).
		\bibitem{Bordcoch2023}
		M.~Bordcoch, C.~N.~Kozameh, and T.~A.~Rojas,
		Phys.\ Rev.\ D \textbf{107}, 104026 (2023).
		\bibitem{Kozameh2026}
		C.~N.~Kozameh and G.~O.~Depaola, in preparation (2026).
		\bibitem{Dyson1952}
		F.~J.~Dyson, Phys.\ Rev.\ \textbf{85}, 631 (1952).
		\bibitem{Britto2005}
		R.~Britto, F.~Cachazo, B.~Feng, and E.~Witten,
		Phys.\ Rev.\ Lett.\ \textbf{94}, 181602 (2005).
		\bibitem{Dominguez1997}
		A.~E.~Dominguez, C.~N.~Kozameh, and M.~Ludvigsen,
		Class.\ Quantum Grav.\ \textbf{14}, 3377 (1997).
		\bibitem{Iriondo1999}
		M.~Iriondo, C.~N.~Kozameh, and A.~Rojas,
		J.\ Math.\ Phys.\ \textbf{40}, 2483 (1999).
	\end{thebibliography}
\end{document}